\documentclass[twocolumn]{aastex631}
\usepackage{amsmath}	
\usepackage{bm}	
\usepackage{gensymb} 

\newcommand{\eppsilon}{$\bm{\varepsilon}$ppsilon}

\submitjournal{ApJ}

\shorttitle{Instrumental Response in EoR PS}
\shortauthors{Barry \& Chokshi}
\graphicspath{{./}{figures/}}

\begin{document}

\title{The Role of the Instrumental Response in 21\,cm EoR Power Spectrum Gridding Analyses}

\author[0000-0003-2064-6979]{Nichole Barry}
\affiliation{International Centre for Radio Astronomy Research \\Curtin University \\Perth, WA 6845, Australia}
\affiliation{The University of Melbourne \\School of Physics \\Parkville, VIC 3010, Australia}
\affiliation{ARC Centre of Excellence for All Sky Astrophysics in 3 Dimensions (ASTRO 3D)}

\author[0000-0003-1130-6390]{Aman Chokshi}
\affiliation{The University of Melbourne \\School of Physics \\Parkville, VIC 3010, Australia}
\affiliation{ARC Centre of Excellence for All Sky Astrophysics in 3 Dimensions (ASTRO 3D)}



\begin{abstract}

Reconstruction of the sky brightness measured by radio interferometers is typically achieved through gridding techniques, or histograms in spatial Fourier space. For Epoch of Reionisation (EoR) 21\,cm power spectrum measurements, extreme levels of gridding resolution are required to reduce spectral contamination, as explored in other works. However, the role of the shape of the Fourier space spreading function, or kernel, also has consequences in reconstructed power spectra. We decompose the instrumental Murchison Widefield Array (MWA) beam into a series of Gaussians and simulate the effects of finite kernel extents and differing shapes in gridding/degridding for optimal map making analyses. For the MWA, we find that the kernel must extend out to 0.001--0.0001\% of the maximum value in order to measure the EoR using foreground avoidance. This requirement changes depending on beam shape, with compact kernels requiring far smaller extents for similar contamination levels at the cost of less-optimal errors. However, simple calibration using pixelated degridding results, regardless of shape of the kernel, cannot recover the EoR due to catastrophic errors caused by the pixel resolution. Including an opaque horizon with widefield beams also causes significant spectral contamination via a beam--horizon interaction that creates an infinitely extended kernel in Fourier space, which cannot be represented well. Thus, our results indicate that simple calibration via degridded models and optimal map making for extreme widefield instrumentation are not feasible.

\end{abstract}





\section{Introduction}

Next-generation telescopes like the Square Kilometre Array (SKA, \citealt{mellema_reionization_2013}) are going to drastically increase the size and sensitivity of available radio interferometers, and therefore increase resolution to reach new astronomical capabilities. However, this leap in technology requires a reevaluation of standard analysis techniques, and whether or not they can achieve the desired scientific goals. In particular, the Epoch of Reionisation (EoR) science case requires extreme levels of spectral precision, and thus the EoR has so far yet remained unseen. 

By using the 21\,cm hyperfine transition of hydrogen, measurements of the EoR during the evolution of the Universe from redshifts 6 through 11 can provide answers to questions regarding cosmology (e.g. \citealt{furlanetto_mesinger_book}) and astrophysics (e.g. \citealt{mirocha_mesinger_book}) via the use of the power spectrum. However, the measurement of the EoR is particularly difficult due to its extremely low brightness ($\sim$10\,mK) when compared to all of the intervening extragalactic and galactic foregrounds ($\sim$300\,K and above) \citep{jelic_foreground_2008}.

Precision analysis of EoR interferometric data sets is inherently difficult due to the measurement basis, which is neither fully harmonic space nor image space. Thus, spectral and spatial harmonic transformations are required depending on the most sensible basis for each particular analysis technique. These transformations depend on the projection, typically either Cartesian or spherical, resulting in Fourier transforms or Spherical Harmonic transforms, respectively (e.g. \citealt{carozzi_imaging_2015}).

Most analyses use a Cartesian projection due to the availability of fast and efficient discrete transforms, namely the Fast Fourier Transform (FFT; \citealt{cooley_algorithm_1965}). Many standard techniques in radio interferometry rely on FFTs, like power spectra analysis and imaging algorithms based on degridding/gridding methods (e.g. \citealt{brouw_aperture,cornwell_noncoplanar_2008}).

In particular, a technique called gridding can be used to reduce data volume whilst providing multiple outputs, such as images and power spectra. Measured visibilities are added to a finely resolved grid in spatial Fourier space, essentially creating a Fourier histogram of the observation. The simplest form of gridding treats the contribution of the visibility as a delta function; however, this leads to aliasing, smearing, and decorrelation. Convolutional gridding avoids these issues by adding the visibility with a chosen spreading function, or kernel, at the location of the baseline in question (see \citealt{offringa_impact_2019} for more details). If the kernel accurately describes the Fourier response of the all-sky instrumental sensitivity, then this plane is the Fourier transform of the orthographic projected sky domain. Only a double spatial-FFT is required to transform from one space to another in this formulism.

Due to the lighter computational load of gridding, many techniques in image and power spectra analysis rely on it, including $w$-projection or $w$-stacking algorithms. However, the gridding kernel needs to be extremely precise for spectral variation studies like the EoR. \citet{offringa_impact_2019} found that to reach the EoR-level precision, overresolution factors of at least 4000 were required for discrete kernels, in addition to over 500 discrete planes to account for $w$-terms. This is difficult in practice, but obtainable with current computing resources. They found that new algorithms which avoid direct gridding, like image domain gridding \citep{van_der_tol_image_2018}, did not require such overresolution factors.

Kernel resolution is undoubtedly important, but there is a broader question on how well the kernel needs to represent the Fourier dual of the instrumental response. Mathematically, errors are considered optimal, or lowest, when the kernel used to grid describes the instrument \citep{tegmark_how_1997,tegmark_how_1997-1}. Furthermore, creating model visibilities for standard calibration techniques requires a well-represented instrumental kernel. In practice, however, it is difficult to represent the kernel accurately (e.g. \citealt{joseph_calibration_2019,chokshi_dual_2021}).

This work explores the limitations with reproducing an accurate instrumental kernel for gridding, both in shape and extent, for EoR power spectrum analyses. We set constraints for kernel extent required in gridding to theoretically measure the EoR, and test if the instrumental response can deviate from the known shape. This is of particular interest to widefield instruments, where extremely low instrumental sensitivity, like secondary sidelobes, can interact with the horizon to produce an infinitely extended kernel in the Cartesian limit. In this particular case, we also explore contamination produced via spatial FFTs and how that propagates to the power spectrum. These results indicate a necessary paradigm shift away from traditional methodologies for the EoR science case, particularly for optimal map making formulisms.

We outline the role of the kernel in gridding (and degridding) in Section~\ref{sec:Instrumental_response_in_gridding_analyses}. In Section~\ref{sec:simulation_framework}, we describe the simulation framework and the control, as well as the quality metric and base level of power spectrum errors within our framework. We show our simulated experiments in Section~\ref{sec:simulated_experiments}, including the role of the kernel extent, systematic propagation into calibration leakage, and effects of the FFT and beam--horizon interactions.


\section{Instrumental response in gridding analyses}
\label{sec:Instrumental_response_in_gridding_analyses} 

A visibility is measured per pair of interferometric elements, where the correlation depends on the separation of the elements in measured wavelengths. If all interferometric elements lie on a 2D plane, then the measurement equation can be expressed as a 2D Fourier transform of the plane-projected sky via the van Cittert–Zernike theorem (see \citealt{carozzi_generalized_2009} for more details). In this Cartesian formalism, it is easiest to express the sky in directional cosines, $\{l,m\}$, and the visibilities in baseline separation in wavelengths, $\{u,v\}$. The measurement equation is then
\begin{equation}
    V(u,v) = \iint \frac{dl\,dm}{\sqrt{1-l^2-m^2}} \,A(l,m) I(l,m) e^{-2 \pi i (ul +vm)},
    \label{eq:grid}
\end{equation}
where $V(u,v)$ are the measured visibilities, $A(l,m)$ is the projected instrumental beam response, and $I(l,m)$ is the specific intensity of the projected sky. 

The visibilities are the 2D Fourier transform of the projected sky, \textit{convolved with} the instrumental beam response. Thus, the complicated response of the instrument is encoded in the visibilities \citep{hamaker_understanding_1996}.  

The formulation described in Equation~\ref{eq:grid} is an over-simplification. There is a third dimension, $w$, which breaks the Fourier transform relation and thus further complicates the measurement equation and the correct beam response. In this work, we only focus on a truly coplanar response via a snapshot in time to investigate the base level of error expected. Therefore, a 2D Fourier transform can be used with coordinate distortions \citep{cornwell_radio-interferometric_1992}. In addition, we further simplify our simulations by assuming there is no frequency dependence on $A(l,m)$ when in reality there is a non-linear, complicated dependence as a function of frequency.

The process of gridding involves placing all visibilities from a frequency/time measurement onto a single, pixelated $uv$-plane to create a $uv$-histogram. However, visibilities will contribute to multiple $uv$-pixels given their baseline locations and the resolutions required to access modes of interest. Thus, visibilities are added to the histogram with a kernel given by $A(u,v)$, or the instrumental response in $uv$-space. This process can be reversed in an action called degridding. Given a $uv$-plane filled with the contribution of model sources, model visibilities can be generated via integration -- again, where the integration kernel used is $A(u,v)$. These model visibilities encode the instrumental response, and thus can be used for calibration or subtraction.

The process of including a kernel is sometimes referred to as \textit{convolutional} gridding/degridding. Conversely, the contribution of a visibility can be treated as a delta function by disregarding the complicated instrumental sensitivity, or, in the specific case of degridding,  incorporating the beam via the apparent intensity $A(l,m) I(l,m) \rightarrow \tilde{I}(l,m)$. However, the discrete nature of the $uv$-plane can result in associated errors, like aliasing, without a kernel.

\subsection{Optimal Map Making}
\label{subsec:OMM}

Optimal map making is a process of lossless data compression which does not destroy cosmological information and produces the smallest (optimal) error bars, initially developed for Cosmic Microwave Background (CMB) data analysis \citep{tegmark_how_1997}. The mathematical formulism of optimal maps has since been adapted for radio interferometry \citep{morales_software_2009}, and is of particular importance to the precise cosmological measurement of the EoR \citep{dillon_mapmaking_2015}. Optimal map making has been used to analyze data from the MWA (i.e. \citealt{dillon_empirical_2015,beardsley_first_2016}), the Hydrogen Epoch of Reionization Array (HERA; i.e. \citealt{the_hera_collaboration_first_2021}), and the Donald C. Backer Precision Array for Probing the Epoch of Reionization (PAPER; i.e. \citealt{ali_paper-64_2015, kolopanis_simplified_2019}).

An important outcome of optimal map making is the ability to propagate errors from the initial measurements through the entirety of the data analysis. Naturally, this means that all operations on the data must be trackable via the variances. This restricts the type of operations that can be used in the optimal map making formulism, and thus has consequences on how we treat the $uv$-plane.

Firstly, the $uv$-plane is only truly optimal when the kernel $A(u,v)$ used in gridding is the actual instrumental response. One can choose to apply an anti-aliasing window function, either directly or as a modification to the instrumental response (i.e. Tapered Gridded Estimator, \citealt{choudhuri_tapering_2016}), to preferentially weight the visibility to the center of the beam at the consequence of higher error bars. While no longer truly optimal, instrumentation with complicated beams can benefit from the trade-off \citep{barry_improving_2019}. However, in order to track this weighting in the propagation of uncertainties, it must be applied in the $A(u,v)$ kernel rather than directly in image space. 

This requirement can be visualised simply by the convolution theorem. If some weighting, $W$, is convolved with the data, $V$, as is the case with gridding, then the image space operation is the multiplication of their Fourier transform equivalents, $\mathfrak{F}(W)$ and $\mathfrak{F}(V)$. Thus, this weighting on the data can be applied in either $uv$-space via a convolution or image space via a multiplication. However, for the propagated variances, $\sigma^2$, the weighting must be applied per $\sigma$, and thus there are two convolved $W$ factors. The same window weighting on the variances in image space is then
\begin{equation}
    (W \ast \sigma) \cdot (W \ast \sigma) \xrightarrow{\mathfrak{F}} \big(\mathfrak{F}(W) \cdot \mathfrak{F}(\sigma)\big) \ast \big(\mathfrak{F}(W) \cdot \mathfrak{F}(\sigma)\big),
    \label{eq:convolve}
\end{equation}
which is notably \textit{not} $\mathfrak{F}(W^2) \cdot \mathfrak{F}(\sigma^2)$. 

Image space corrections that are typical of radio astronomy, like using the conjunction of image space padding with window functions to remove unstable edges (see \citealt{van_der_tol_image_2018} for more details), cannot be used when propagating variances in an optimal map making formulism. In this work, we only apply methodologies that allow for the ability to propagate variances from the measurement, and thus we are restricted in the operations that can be used.




\section{Simulation framework}
\label{sec:simulation_framework} 

To investigate the effects of the instrumental response in gridding for an optimal map making framework, we employ an in situ simulation technique using a data analysis pipeline. This allows the ability to simulate a real analysis response with all the potential contamination effects whilst maintaining control over inputs and experimental variations. This framework was first introduced in this pipeline by \citet{barry_calibration_2016}, which goes into greater detail. In this section, we summarise the key points necessary to understand the experimental procedure used in Section~\ref{sec:simulated_experiments}.

\subsection{Analysis pipeline source}
\label{subsec:analysis_pipeline_source} 

The analysis framework for our experiment is an open-source, EoR-specific pipeline composed of 1) Fast Holographic Deconvolution\footnote{\url{https://github.com/EoRImaging/FHD}} (FHD; \citealt{sullivan_fast_2012,barry_fhd/eppsilon_2019}) and 2) Error Propagated Power Spectra with Interleaved Observed Noise\footnote{\url{https://github.com/EoRImaging/eppsilon}} ({\eppsilon}; \citealt{barry_fhd/eppsilon_2019}), hereafter called FHD/{\eppsilon}.

FHD/{\eppsilon} has been used to generate competitive EoR upper limits \citep{beardsley_first_2016, barry_improving_2019, li_first_2019}, investigate calibration systematics \citep{barry_calibration_2016,li_comparing_2018, byrne_fundamental_2019, zhang_impact_2020}, and probe subtleties in foreground and RFI subtraction \citep{pober_importance_2016,carroll_KGS, kerrigan_improved_2018, wilensky_absolving_2019, wilensky_quantifying_2020,byrne_map_2021}. This analysis pipeline has been proven to be able to quantitatively characterise instrumental effects and spectral precision requirements for EoR science. 

Of course, how these effects propagate to the final measurement space depends on the specific analysis techniques used. For the purposes of our simulations on single observations, these can be summarised chronologically as the following steps:
\begin{description}
    \item[\textbf{Beam kernel}] We generate a hyper-resolved table of the spatial Fourier transform $\{u,v\}$ from the instrumental beam response on the sky in directional-cosine space $\{l,m\}$. This flattens the 3D sky response of the instrument into a 2D Fourier representation via orthogonal projection, and is the estimation for the instrumental contribution to visibility measurements. For EoR analysis, this is generally calculated at a high hyper-resolution \citep{offringa_precision_2019}. We reduce resolution errors further by performing bilinear interpolation of the hyper-resolved table during analysis.
    \item[\textbf{Point-source model}] Each source's celestial position and flux density is analytically Fourier transformed to a $uv$-grid. All of the sources' $uv$-grids are added together to generate a single, discrete model $uv$-plane in the absence of instrumental effects \citep{kerrigan_improved_2018}. Since there are commonly over 50,000 sources per field, we hold the sky fixed in order to only generate one model $uv$-plane per observation, and correct for a mean spectral index after model visbility estimation.
    \item[\textbf{Model visibility estimation}] We perform degridding of the singular model $uv$-plane to create model visibilities. In brief, we multiply the hyper-resolved beam kernel with the model $uv$-plane at each baseline location and integrate over all pixels per baseline. Since we held the sky fixed, we must then move each baseline for each time and frequency bin and re-integrate. The response of the instrument is encoded via the beam kernel, and represents an estimation for how the instrument measures the point-source sky. 
    \item[\textbf{Calibration}] The gains of each individual station/tile is found via a linear least squares reduction of a system of equations between the data visibilities and the newly made model visibilities \citep{mitchell_real-time_2008} for baselines longer than 50\,wavelengths \citep{patil_systematic_2016}. We do not constrain spectral smoothness in this work in order to show the base level of error that must be mitigated. 
    \item[\textbf{Discrete $\bm{uv}$-plane estimation}] We then grid each data/model visibility to a $uv$-plane per frequency. This operation is performed by multiplying the hyper-resolved beam kernel with the visibility value and adding the pixelated result to the $uv$-plane. This is analogous to a histogram of the data, where the spread of the visibility in the $uv$-plane is encoded by the beam kernel. 
    \item[\textbf{Time-interleaved sets}] The $uv$-plane is split into two sets which are interleaved in time. This choice allows us to be able to estimate the cross-power spectrum via sums and differences in addition to other diagnostics \citep{barry_fhd/eppsilon_2019}.
    \item[\textbf{Residual $\bm{uv}$-plane}] Optionally, we subtract the model $uv$-plane from the data $uv$-plane for each frequency to generate a residual $uv$-plane.
    \item[\textbf{Frequency Lomb-Scargle transform}] We apply a low-sidelobe window function along each spatial $uv$-pixel and perform a spectral Lomb-Scargle transform to go to full harmonic space. The Lomb-Scargle periodogram ensures that we capture any power that is hidden in the covariances due to uneven sampling \citep{lomb_least-squares_1976,scargle_studies_1982}.
    \item[\textbf{Cross-power estimation and binning}] Finally, we combine the sums and differences of the interleaved time sets in $k$-space to form the cross-power spectrum and bin to either cylindrical or spherical power spectra.
\end{description}
This is a simplified set of modules; we did not perform image processing with our single, short-time interval observation simulations in order to reduce the complexity of the analysis. 
Our simulations use a MWA Phase I layout \citep{tingay_murchison_2013}, which has a well-covered central core in $uv$-space with slowly decreasing coverage after 50\,wavelengths. 

\subsection{Quality metric}
\label{subsec:quality_metrics}

Our metric for statistical 21\,cm EoR detection in this work is the power spectrum. By harnessing the ability to prescribe an almost exact redshift to the narrow line emission of the 21\,cm hyperfine transition of Hydrogen, we can form measurements of the power for each mode on the sky as a function of $z$.

In this work, we use a \textit{reconstructed power spectrum} as described by \citet{morales_understanding_2019}.  This entails binning the sky image into a fully perpendicular cube in comoving megaparsecs along the spatial \textit{and} frequency dimensions. Other forms of power spectra estimation (i.e. delay spectra) and other statistical measurements (i.e. the bispectrum) experience contamination differently, and are not explored in this work.

We bypass creating images and instead directly form a $\{u,v,f\}$-plane of our simulated data, which only requires a simple coordinate transform to create a $\{k_x,k_y,f\}$-plane. A further frequency transform (specifically in the case of FHD/{\eppsilon}, a Lomb-Scargle periodogram) along with another simple coordinate transform creates a fully perpendicular $k$-cube \citep{morales_toward_2004}. An antialiasing window is used during the frequency transform, which reduces unwanted systematics at the cost of a smaller effective bandwidth. Using the assumption of isotropy and homogeneity of the EoR, we average the $k$-cube in spherical shells to create 1D power spectra as a function of $|k|$. 

\begin{figure}
\centering
	\includegraphics[width =\columnwidth]{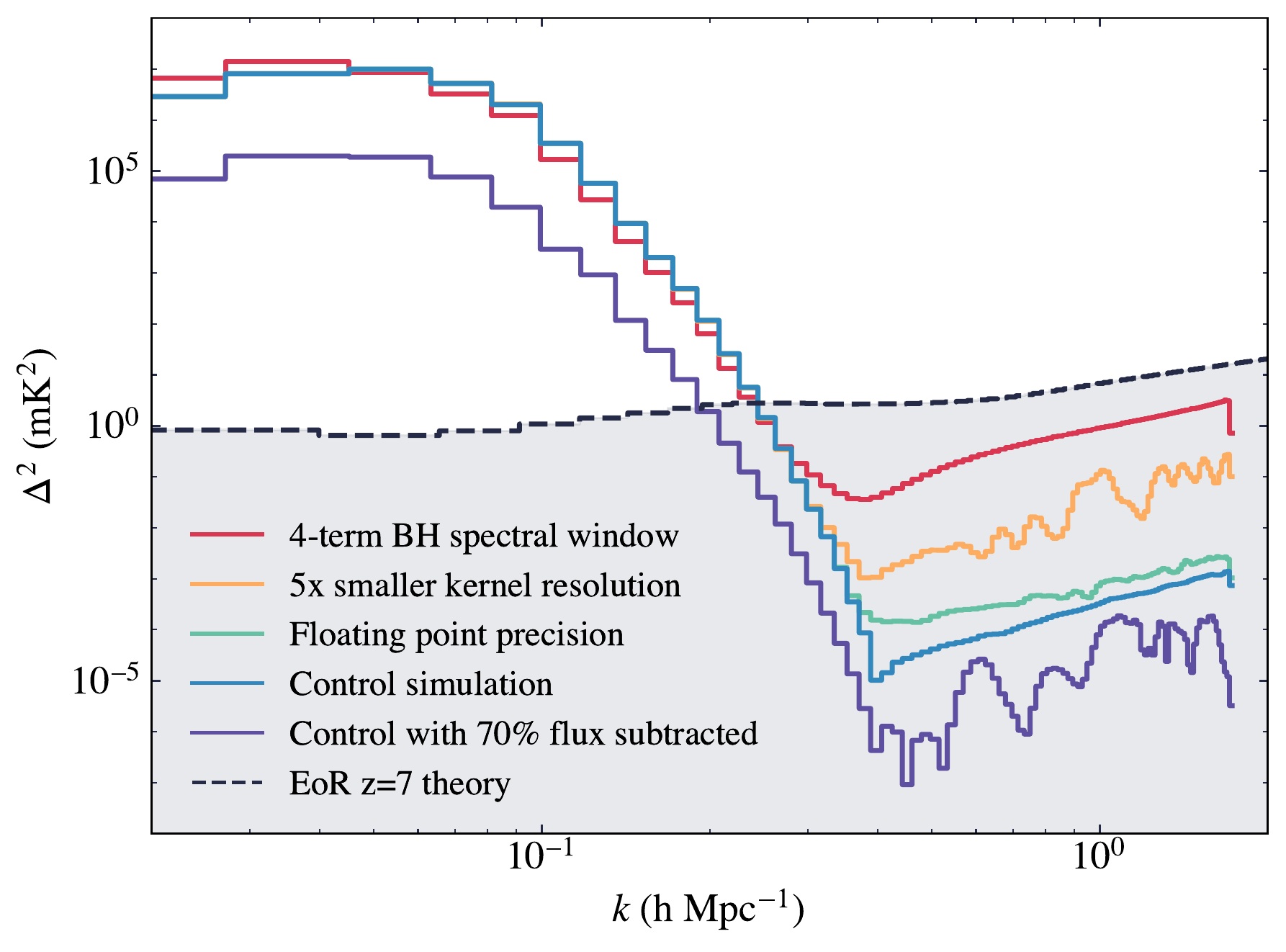}
	\caption{An example of the base level of errors expected in our simulation framework compared to a fiducial EoR theory (dashed) and the control simulation (blue). These systematics include changing specific factors in the control, such as changing the 7-term Blackman-Harris (BH) $f$-to-$k$ window to a 4-term BH (red), reducing the $uv$-kernel resolution by five times (yellow), and using floating point precision in the gridding step instead of double point precision (green). We also show a realistic foreground subtraction of the control (purple). Errors must be within the shaded region in order to theoretically detect the EoR.}
	\label{fig:control}
\end{figure}

Low $k$-modes will be contaminated by the chromatic response of the instrument \citep{datta_bright_2010,morales_four_2012,parsons_per-baseline_2012,trott_impact_2012,vedantham_imaging_2012,hazelton_fundamental_2013,pober_opening_2013,thyagarajan_study_2013,liu_epoch_2014}. Higher $k$-modes are theoretically free from this contamination, and thus this region, known as the EoR window, holds the modes of interest for foreground avoidance techniques. Since our simulations are noise-less and without an underlying EoR signal, any remaining power in high $k$-modes indicate analysis-based systematics. Figure~\ref{fig:control} shows the 1D power spectrum for the simulations described in Section~\ref{subsec:in_situ_simulation} along with a theoretical realisation of the EoR from \citet{barry_improving_2019} in units of mK$^2$. These simulations have varying degrees of error, but all could theoretically measure the 21\,cm signal in the EoR window since their level of error is below the expected power. 

\subsection{In situ simulation}
\label{subsec:in_situ_simulation} 

The FHD/{\eppsilon} pipeline was built to analyse observations, but can also be harnessed as an in situ instrumental simulation to test for various spectral precision effects.

By default, FHD/{\eppsilon} runs an instrumental simulation with point-source foregrounds for every analysis run -- this constitutes the model visibilities. These model visibilities encode the instrument via the beam kernel, and are made with typically 50,000 known sources or more. We can use these model visibilities as an input into the pipeline to perform an ad hoc, in situ simulation.

The now-input visibilities are treated the same as data; all of the steps described in Section~\ref{subsec:analysis_pipeline_source} are performed, and new model visibilities are made alongside our input visibilities. By varying the amount of known point-sources in the new model, we can simulate our lack of knowledge of the flux density of the sky. 

More importantly, we can simulate our lack of knowledge about the instrumental beam response. This will form the basis of our experiments in Section~\ref{sec:simulated_experiments}. We can create the input simulated visibilities using a specific beam kernel, and then during the in situ simulation, use a different, modified beam kernel in analysis. Errors that arise from using a gridding kernel that is different than the encoded instrumental response will then become apparent in the power spectra. When compared to a control run, we can then quantify the excess error.

\subsubsection{The Gaussian decomposition and semi-analytic input}
\label{subsubsec:the_input}

The input model visibilities are treated as our simulated data in the FHD/{\eppsilon} in situ simulation. These were created from a model $uv$-plane of an EoR0 target field (centred around RA=0\degree, Dec=-27\degree) zenith observation with 49,572 point sources from the GLEAM catalogue \citep{hurley-walker_GLEAM}. There is no thermal noise, no ionospheric effects, and no diffuse emission.

The choice of instrumental beam kernel when degridding this model $uv$-plane will determine the encoded instrumental response in the input visibilities. Thus, we want to ensure that we are including as few known systematics as possible. We create a new representation of the beam for this purpose -- the Gaussian decomposition of the instrumental sky response and its analytic Fourier transformation.

We create a hyper-resolved directional cosine image from outputs of a FEKO simulation of an averaged embedded element beam with mutual coupling \citep{sutinjo_understanding_2015}. This is a 2D representation of the sky response, and is the Fourier dual of our desired $uv$-space instrumental kernel. However, instead of performing FFTs to transform the instrumental model, we fit a series of Gaussians to the image to create an analytic decomposition of the instrumental sky response. Thus, the Fourier transform is fully analytic and calculable. 

The sum of analytic Fourier transforms of each Gaussian produces an alias-free $uv$-kernel. In addition, we can calculate the exact beam values at every pixel for each unique baseline location using the analytic decomposition function, rather than relying on interpolation or nearest-neighbor methods to calculate the beam values per pixel from a discrete kernel. We use this to reduce gridding-resolution effects, in addition to calculating beam values for all included pixels above machine-level precision. This is a semi-analytic formulism; the visibilities are analytic in every way \textit{except} that the $uv$-plane to be degridded is inherently pixelated. Unfortunately, this semi-analytic formulism is not a realistic option for large-scale data analyses for extreme widefield instruments, and thus we only use this for our input. 

Other functions can be chosen in place of Gaussians as long as they are analytic, such as prolate spheriodals, Bessel functions, or spherical harmonics. The main reason for choosing Gaussians with our simulation is that they are extremely simple whilst still achieving the desired end result. In the future, finding an analytic form of the projection of spherical harmonics to 2D Cartesian $uv$-space would avoid fitting procedures since FEKO simulations can output spherical harmonics. However, this is non-trivial, potentially very time-consuming, and adds unnecessary complexity to our basic simulations. 

The Gaussian decomposition of the instrumental sky response is composed of 18 individual Gaussians, and fits the FEKO simulation to within 0.3\%. Nevertheless, one important feature is excluded: the horizon. Gaussians are infinite and thus the Gaussian decomposition of the beam has no horizon. The effect of including a horizon is explored in Section~\ref{subsec:The_effects_of_the_horizon}. 

\subsubsection{The discrete control and associated errors}
\label{subsubsec:the_control}

The control run determines the level of base error that is expected when the 2D Gaussian decomposition of the beam is used throughout the analysis. Whatever systematics that are present in the control will also be present in the experiments. To first order, we can assume these systematics are additive in the experiment. Given our simplified approach (i.e. no imaging with a smooth $uv$-coverage and a perfectly 2D array), this basic assumption should hold.

For our control, we use a representative gridding resolution, which will result in a base level of error. For these experiments, we use a 250x hyper-resolved beam with bilinear interpolation to pixel centers, resulting in the control simulation (blue) in Figure~\ref{fig:control}. If we were to decrease the resolution by five times (yellow, Figure~\ref{fig:control}), however, we would still be below the level of the EoR.

Throughout the analysis, we perform calculations at double precision. If the gridding operation is performed at floating point precision (green, Figure~\ref{fig:control}), this would have resulted in a small increase in error. We also use a 7-term Blackman-Harris spectral window function in transforming from $\{k_x,k_y,f\}$ to $\{k_x,k_y,k_z\}$ due to the modest increase in the base level of errors with a 4-term Blackman-Harris spectral window function (red, Figure~\ref{fig:control}).

We can optionally allow the in situ simulation to subtract 11,188 of the brightest sources in the sky as seen by the instrument, or roughly 70\% of the apparent flux density (purple, Figure~\ref{fig:control}). This represents a realistic subtraction of flux density on the sky that we can expect to achieve in real data analysis. In addition, by only subtracting a subset of sources, we enable a realistic dynamic range problem. Errors that arise in the analysis that couple the foregrounds to the measurement modes are only observable when foregrounds or foreground residuals are present. As seen in Figure~\ref{fig:control}, we finally reach an unknown, varying systematic with the foreground subtraction that is six orders of magnitude below the expected EoR power.

The resulting error in the measurement modes of interest is the base level expected in Section~\ref{subsec:Beam_kernel_extent} and Section~\ref{subsec:calibration_bias} (blue for unsubtracted power, purple for subtracted power, Figure~\ref{fig:control}). For Section~\ref{subsec:The_effects_of_the_horizon}, we relax the hyperresolution parameters for computational reasons (yellow, Figure~\ref{fig:control}). Any additional errors can be quantified as resulting from the specific variable changes in our analysis.


\section{Simulated experiments}
\label{sec:simulated_experiments}

We present three simulated experiments to investigate the effects of the gridding/degridding kernel on the EoR power spectrum, namely the choice of gridding kernel in optimal map making (Section~\ref{subsec:Beam_kernel_extent}), calibration bias (Section~\ref{subsec:calibration_bias}), and inclusion of the horizon (Section~\ref{subsec:The_effects_of_the_horizon}). 

\begin{figure*}
\centering
	\includegraphics[width =\textwidth]{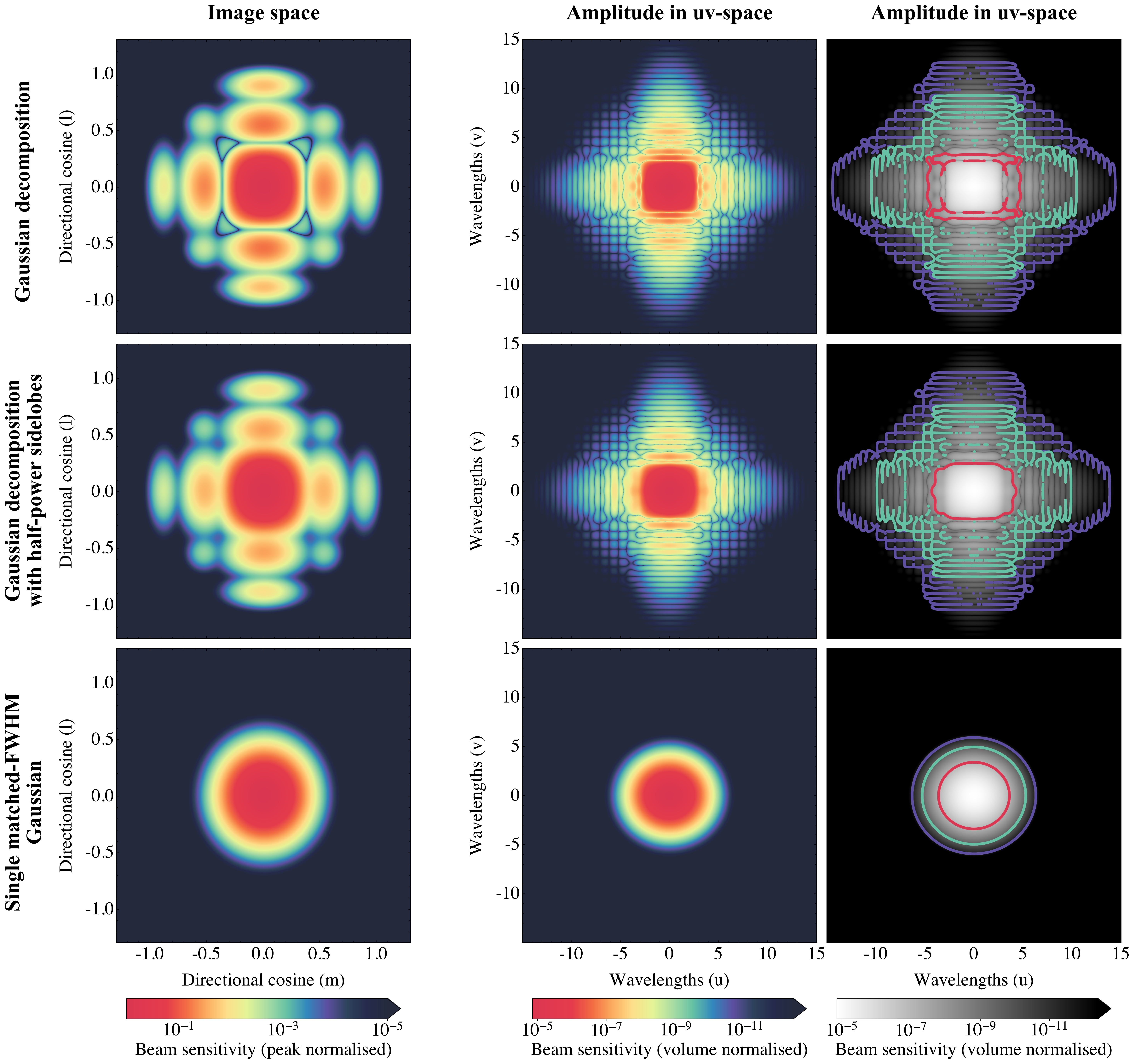}
	\caption{The three types of projected image-space beam shapes (left column) used in this work and their respective analytic 2D Fourier transforms (amplitude only, middle column) along with derivative-matched extents (right column; red: 0.01\%, green: 0.0001\%, purple: 0.000001\% of the Gaussian decomposition maximum). The top row is the Gaussian decomposition of the instrumental MWA beam, accurate to within 0.3\%, which is used as our control in our experiments. The middle row is a Gaussian decomposition where all components except the primary lobe component have been reduced by half. The bottom row is a single Gaussian with a FWHM matched to the instrumental beam. These kernels represent a perfect instrumental beam, an imperfect instrumental beam, and a non-instrumental beam, respectively. Fewer pixels are required for comparable levels of power spectrum contamination for the non-instrumental beam. }
	\label{fig:panel}
\end{figure*}

Each experiment uses three types of $uv$-kernels: the Gaussian decomposition of the beam (top row, Figure~\ref{fig:panel}), the Gaussian decomposition of the beam with all components at half power except the primary lobe component (second row, Figure~\ref{fig:panel}), and a single Gaussian matched to the FWHM of the primary lobe (bottom row, Figure~\ref{fig:panel}). These kernels represent a perfect instrumental beam, an imperfect instrumental beam, and a non-instrumental beam, respectively. As seen in Figure~\ref{fig:panel}, the presence of sidelobes causes an interference pattern in the analytic Fourier transform. 

We only show the E--W instrumental amplitude response for brevity. 


\subsection{Gridding kernels in optimal map making}
\label{subsec:Beam_kernel_extent}

The $uv$-kernel is infinite in extent yet highly peaked, and therefore a choice must be made in producing a finite representation. This choice is influenced by efficiency requirements; the larger the kernel extent, the more pixels involved in operations that are repeated millions of times, and therefore a compact yet accurate representation is preferred. 

However, choosing a finite extent can create a sharp discontinuity in the $uv$-kernel at the boundary. As baselines move as a function of frequency, the pixels in the $uv$-plane that contribute will change. Thus, if there exists a sharp discontinuity in the $uv$-kernel, there will be abrupt transitions as a function of frequency, which will propagate to the power spectrum as a systematic.

A beam kernel cut is equivalent to convolving a sinc-like function in image space. This will undoubtedly generate contamination in the power spectrum. However, our options are limited. We cannot apply image-space corrections easily, a common mitigation technique, due to our choice of optimal map making (see Section~\ref{subsec:OMM}). In addition, we cannot realistically grid all pixels for every single baseline given the extreme widefield nature of the MWA sidelobes. Therefore, the best option for optimal map making is to find a level of contamination caused by the extent of the beam kernel that is sufficiently below the power level of the EoR.

This can be partially mitigated if the extent is chosen to be a contour. The beam value of the contour can be subtracted off the entire kernel to reduce the step to zero, and then the $uv$-kernel can be renormalised. This deforms the $uv$-kernel slightly, making it less instrumental at the cost of reducing systematics. Nevertheless, this technique will not change the fact that the discontinuity remains. We apply this mitigation method to show the minimum level of error that is expected. The effects of the small deformation via the renomalisation appear minimal, as will become evident.

We choose three different extents to highlight the dependence on the edge discontinuity, as shown in third column of Figure~\ref{fig:panel}. For the Gaussian decomposition (top right), these occur at contours of 0.01\% (red), 0.0001\% (green), and 0.000001\% (purple) of the maximum $uv$-kernel value,  corresponding to approximately 10\,$\lambda$, 20\,$\lambda$, and 30\,$\lambda$ extents in the $u$ direction, respectively. In order to reproduce similarly behaved responses in the other beam kernel types, we calculate the mean derivative along the subtracted contour edge in the Gaussian decomposition kernel and then find the corresponding contour value for the kernel in question. The edge derivatives are therefore roughly matched between the different kernel shapes. 


Matched-derivative contours result in vastly different kernel extents. The single, matched-FWHM Gaussian (bottom right, Figure~\ref{fig:panel}) has a very compact kernel where the maximum extent is less than 15\,$\lambda$. This is extremely useful in gridding/degridding analyses, as the number of pixels required per visibility is four times fewer. In contrast, the kernel with half-power sidelobes (middle right, Figure~\ref{fig:panel}) has not changed dramatically in extent, only decreasing by a few wavelengths. This indicates that it is difficult to reduce the contribution of sidelobes to the $uv$-kernel extent, and that even a 50\% decrease in sidelobe contribution has very little effect. 

The kernel extent due to widefield sidelobes has consequences in instrumental design. Significantly spaced sidelobes create an interference pattern in Fourier space, which extends the contribution of the kernel out much further than if those sidelobes were not present. Decreasing the sidelobes by 50\%, a technical feat in-of-itself when designing an instrument, would not significantly change this extent. However, if sidelobes were able to be sufficiently suppressed, $\mathcal{O}(N^2)$ analyses could expect a speed-up factor of sixteen given the reduction of pixels per visibility required in analysis.

\begin{figure}
\centering
	\includegraphics[width =\columnwidth]{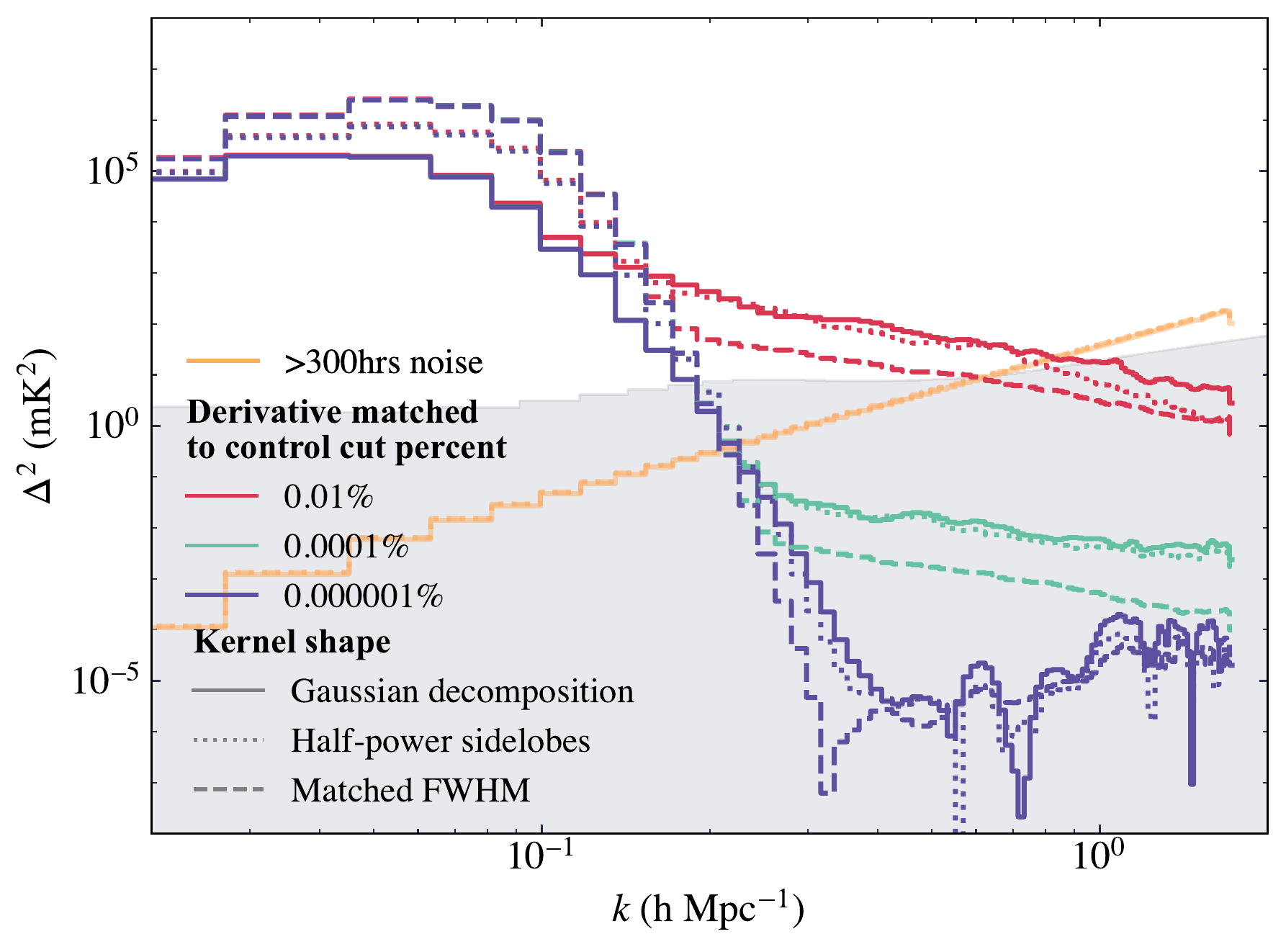}
	\caption{The 1D power for the various extents and shapes with 70\% of the flux density subtracted to be representative of measurements. The extents of the Gaussian decomposition (solid lines) are 0.01\% (red), 0.0001\% (green), and 0.000001\% (purple) of the maximum beam value. For the half-power sidelobe beam (dotted lines) and FWHM-matched Gaussian beam (dashed lines), the contours correspond approximately the same mean edge-derivative as the Gaussian decomposition as shown in Figure~\ref{fig:panel}. Power must be within the shaded region in order to theoretically detect the EoR. Thermal noise (1$\sigma$, yellow) for 333\,hrs of data for each beam shape shows there is very little degradation of optimal uncertainties.}
	\label{fig:stamps_power_abs}
\end{figure}

The resulting power spectra from each $uv$-kernel shape and extent are shown in Figure~\ref{fig:stamps_power_abs}. In each case, the input simulation was calculated with a Gaussian decomposition beam in our semi-analytic formulism described in Section~\ref{subsubsec:the_input}. We have removed 70\% of the flux density through a degridded model. These power spectra are without calibration errors, and therefore only probe the effect of the $uv$-kernel specified in gridding/degridding.

The high $k$-modes, or EoR window, decrease in power as the extent is increased for each kernel shape. The Gaussian decomposition kernel and the kernel with half-power sidelobes decrease at roughly the same rate given a derivative-matched contour, whereas the matched-FWHM Gaussian kernel decreases quicker. This can be attributed to, in part, the differences in the derivative around the contour in the Gaussian decomposition kernel, which vary and are not fully captured by the mean. All kernel shapes could theoretically detect the EoR with a derivative-matched contour of $\sim$0.0001\% for a majority of EoR window modes.

The low $k$-modes, however, indicate the importance of faithfully reconstructing the beam in foreground removal. The Gaussian decomposition kernel reduced the foregrounds more effectively than the other kernel shapes because it better represented the input, and thus has lower power on low $k$-modes. Nevertheless, there is still a significant amount of power left in this region because the Gaussian decomposition kernel used did not \textit{perfectly} match the input, which was semi-analytic in nature. This may indicate optimal map making analyses that use pixelated gridding/degridding procedures will be unable to remove enough power to measure the EoR on low $k$-modes without significantly finer resolution $uv$-planes. Additional procedures, like optimal map making with covariance weighting (e.g. \citealt{ali_paper-64_2015}) could potentially remove enough power, but may suffer from signal loss \citep{cheng_characterizing_2018}.

Misrepresenting the beam to reduce systematics at high $k$-modes (for example, via the single matched-FWHM Gaussian beam) essentially results in a convolutional image-space weighting \citep{choudhuri_tapering_2016}. This may result in non-optimal errors \citep{tegmark_how_1997, tegmark_how_1997-1}, and no longer gives the orthographic projection of the sky. Nevertheless, it reduces extent-based systematics and requires fewer pixels during computation.

We highlight the change in the 1$\sigma$ thermal noise level due to the various beam kernel shapes (yellow, Figure~\ref{fig:stamps_power_abs}) given 333\,hrs of data. Even though the beam kernel with half-power sidelobes  and the matched-FWHM Gaussian kernel produce non-optimal errors, their difference from the optimal errors of the Gaussian decomposition beam kernel is negligible. While the beam shapes are quite different from one another past $\sim$$15\,\lambda$, the majority of the beam volume comes from the center of the beam where they are almost identical. Thus, the 1$\sigma$ thermal noise level for the matched-FWHM Gaussian kernel is only 8\% higher than optimal. 

Out of the three kernel shapes, the matched-FWHM Gaussian kernel is preferred during the gridding process due to its ability to reduce systematics whilst requiring fewer pixels during computation. This does come at the cost of non-optimal errors, but the change is almost negligible. While the framework of optimal map making is extremely useful for reconstructing maps with fully propagated uncertainties, it is more practical to be non-optimal to reduce widefield sidelobe effects in the analysis load. 

\subsection{Calibration errors with misrepresentations of the instrument}
\label{subsec:calibration_bias}

Given the apparent success of compact, non-instrumental $uv$-kernels in the gridding procedure of Section~\ref{subsec:Beam_kernel_extent}, it may be tempting to use a similar $uv$-kernel to generate model visibilities for calibration through degridding. However, using a non-instrumental beam is equivalent to a weighting in image space, and thus the model visibilities would no longer represent the instrument, leading to an inaccurate calibration.

In order to quantify the resulting errors that occur in calibration due to misrepresentations of the instrument, we now modify the beam kernel used for degridding. Again, the simulated data visibilities are calculated with a Gaussian decomposition beam in our semi-analytic formulism in Section~\ref{subsubsec:the_input}. For this experiment, we will also use \textit{all} of the flux density on the sky in calibration to bypass errors caused from frequency-dependent point spread functions of unmodelled sources \citep{barry_calibration_2016}. Any spectral structure in the calibration will thus be solely due to misrepresentations of the beam kernel.

We use a simple calibration procedure: a linear-least squares solution between the data and model visibilities to get the resulting gain parameters of each interferometric element per frequency \citep{mitchell_real-time_2008}. In general, this represents the most basic form of calibration and will give us an idea of systematics that will then need to be mitigated through more sophisticated processes like frequency regularisation. 

\begin{figure}
\centering
	\includegraphics[width =\columnwidth]{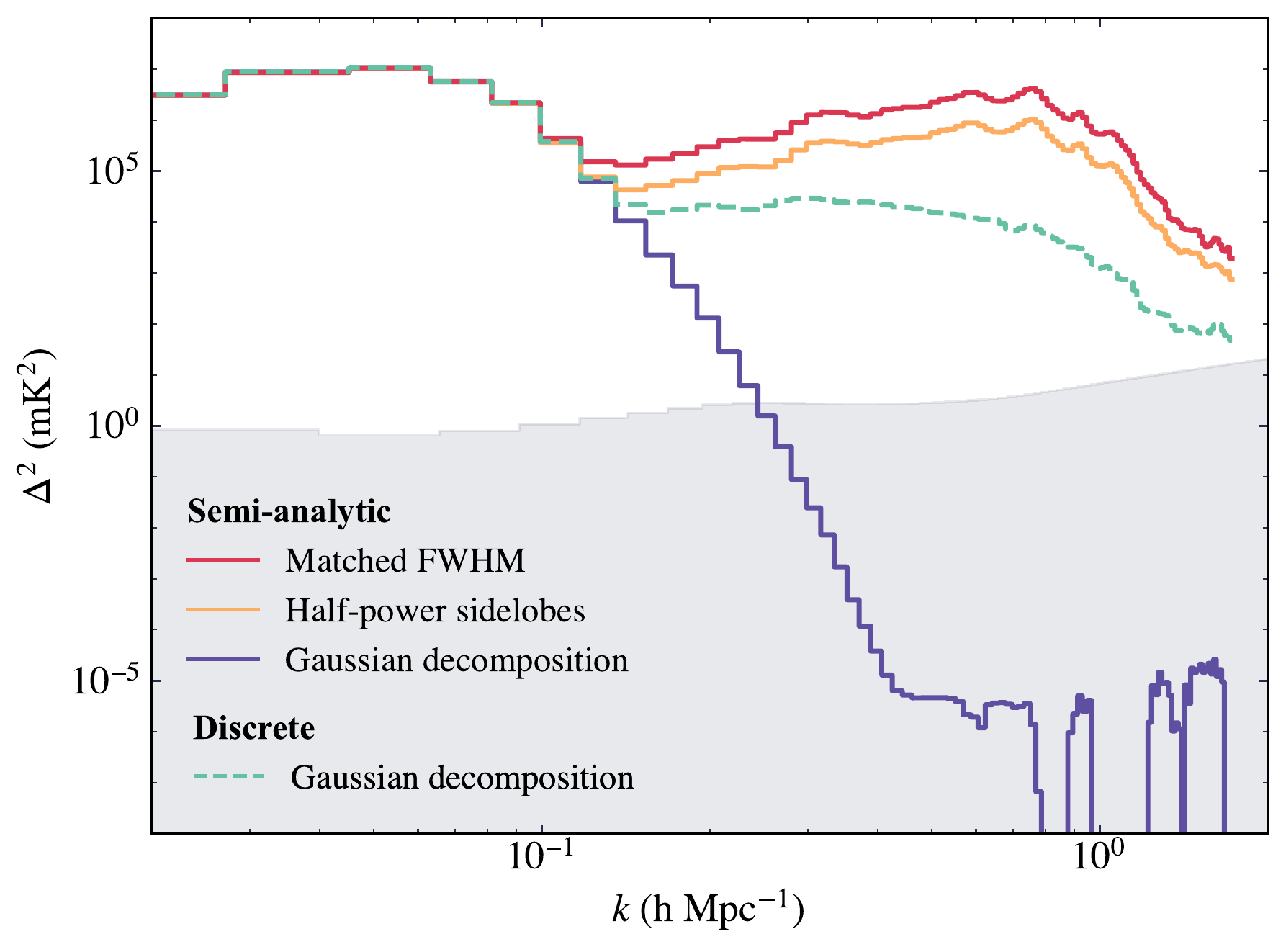}
	\caption{Calibration systematic errors caused by misrepresentations of the control input beam in degridding. Using the semi-analytic Gaussian decomposition beam (see Section~\ref{subsubsec:the_input}) with all the flux density on the sky modelled perfectly results in little-to-no calibration errors in the EoR window as expected (purple). Modifying the degridding kernel to the semi-analytic half-power sidelobe beam (yellow) or the semi-analytic FWHM-matched Gaussian beam (red) results in catastrophic calibration errors. However, even the typical, discrete Gaussian decomposition has contamination the precludes an EoR measurement (dashed green).}
	\label{fig:cal_ps}
\end{figure}

As a control, we first test the resulting errors in calibration when the semi-analytic Gaussian decomposition is used. Again, the semi-analytic formulism uses the exact analytic Fourier transform of the beam kernel for all contributing pixels rather than the hyperresolved lookup table. As expected, there are little to no errors present in calibration, and there are many accessible high $k$-modes (purple, Figure~\ref{fig:cal_ps}). No flux was subtracted in this test or any of the following experiments in order to compare to this base level of error; otherwise, there is essentially no power left in the control since everything subtracts almost perfectly. 

We now modify the degridding beam shape to the semi-analytic beam kernel where all Gaussian components except the main lobe have been reduced by half (yellow, Figure~\ref{fig:cal_ps}). All $k$-modes are contaminated by calibration errors, and the EoR cannot be measured. This is unfortunate -- this means that in order for a simple calibration to work for EoR science, even the sidelobes of the instrumental beam must be represented well. 

Modifying the degridding beam shape to the semi-analytic single FWHM-matched Gaussian creates even more calibration error (red, Figure~\ref{fig:cal_ps}). When compared to the semi-analytic half-power sidelobe kernel, this error is between 50\% and 200\% more at high $k$-modes. Even though this kernel was preferred in gridding within an optimal map making framework for its suppression of EoR window spectral contamination and for its compactness, it cannot be used for degridding. This again reiterates that calibration is an instrumental procedure and thus the degridding kernel used must represent the instrument. 




However, a further complication arises when the degridding procedure uses our typical hyperresolved beam like in Section~\ref{subsec:Beam_kernel_extent} instead of the semi-analytic calculation. This fully discrete formulism results in EoR-precluding calibration errors, even with the Gaussian decomposition beam kernel (dashed green, Figure~\ref{fig:cal_ps}). This hyperresolved beam kernel was at the same resolution as those that were sufficient for gridding. It is apparent from this contamination that the requirements stated in \citet{offringa_impact_2019} for gridding are not necessarily the same as the requirements for degridding calibration schemes.

More sophisticated calibration schemes will of course do better at reducing spectral contamination into high $k$-modes. This will depend on the level of flux density modelled, the refinement of the frequency regularisation, and the complexity of the instrument in question. For example, if the frequency dependence of an instrument can be accurately represented by a low-order polynomial fit, then most spectral contamination in the EoR window can be avoided \citep{barry_calibration_2016}.

Nevertheless, our experiments shown in Figure~\ref{fig:cal_ps} do not bode well for degridding calibration schemes. Representing the instrumental kernel well enough will prove difficult, especially given its complex and time-varying nature \citep{joseph_calibration_2019,chokshi_dual_2021}. Even if we could capture its response, we would have to greatly increase the resolutions of degridding to levels that we may not be able to achieve. This suggests a paradigm shift in our model generation and calibration procedures.

Thankfully, model visibility generation is not beholden to optimal map making and being able to propagate errors. Thus, degridding does not necessarily have to be used -- as long as the result is instrumental model visibilities, any procedure will work in theory. In addition, new calibration schemes are helping to remove the dependence on model visibilities altogether. (cite everyone really).




\subsection{The effects of the horizon}
\label{subsec:The_effects_of_the_horizon}
One important aspect of the gridding kernel that has been omitted in this work thus far has been the horizon. Its inclusion into our analytic framework is possible, yet unsustainable on a case-by-case basis for real observational data given that it requires a pixel-by-pixel convolution. Nevertheless, it reveals detriments for wide-field telescopes and traditional optimal map making analysis techniques. 

\begin{figure*}
\centering
	\includegraphics[width =\textwidth]{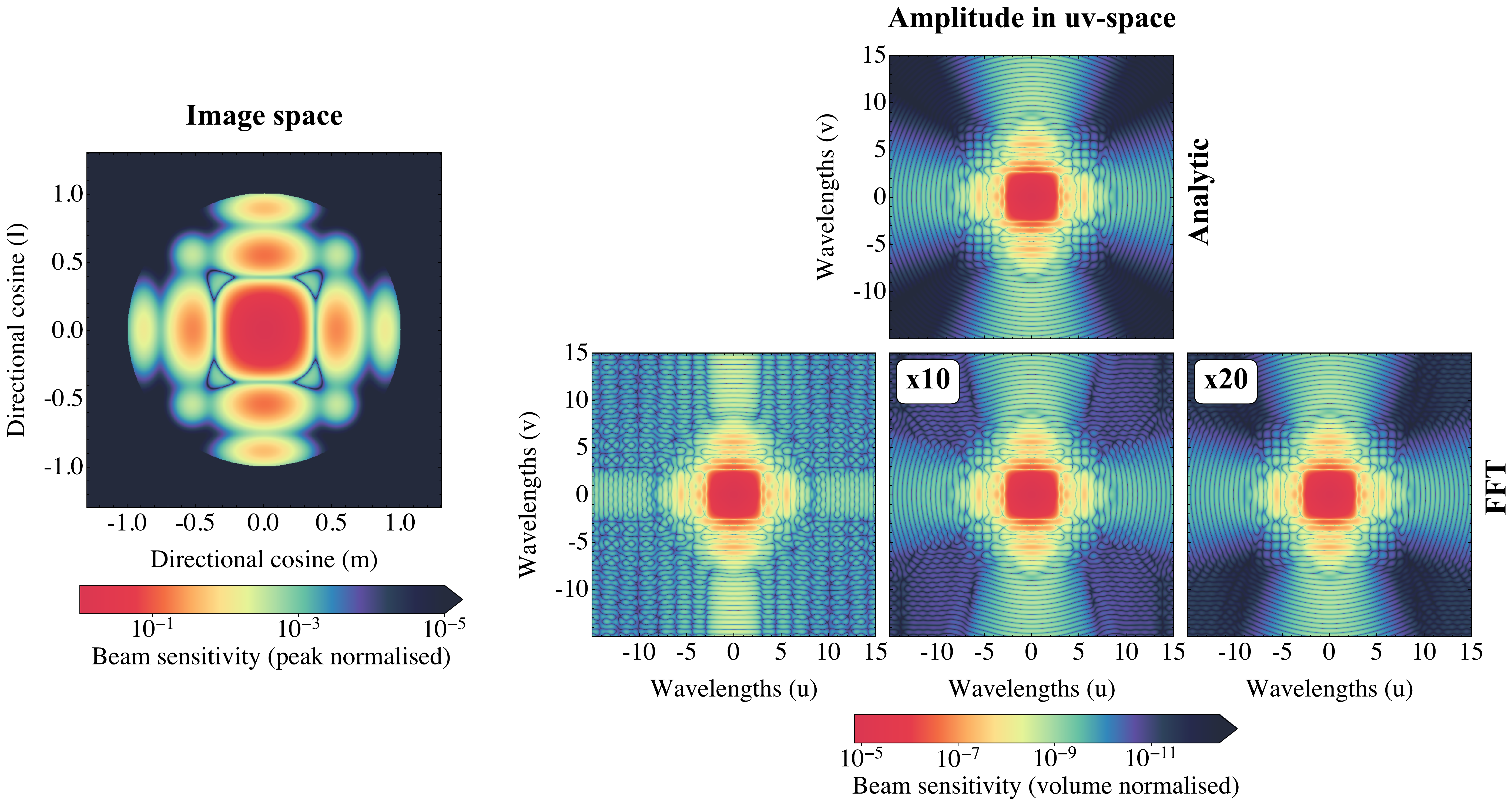}
	\caption{The image space (left panel) and $uv$-space amplitude (right panels) of the Gaussian decomposition with the horizon. The $uv$-space amplitude can be calculated analytically with a convolution (top right panel) or calculated via FFTs (bottom right row). FFT aliasing depends on the image-space resolution, and we show the base resolution (left in row), 10x resolution (middle in row), and 20x resolution (right in row) to highlight the reduction of aliasing artifacts towards higher resolutions.}
	\label{fig:fft_panel}
\end{figure*}

In flat $lm$-space, the horizon is a circular step function with values of 1 inside a radius of 1, and with values of 0 elsewhere. We multiply this by the beam kernel formulations to generate more realistic representations, such as the left panel of Figure~\ref{fig:fft_panel}. Thus, the analytic Fourier transform is the convolution of the Gaussian decomposition and the circular step function in Fourier space (top panel, Figure~\ref{fig:fft_panel}). We reduce the kernel resolution by 5x in order to perform the analytic convolution in a reasonable time, which increases the base level of expected error (from blue to yellow, Figure~\ref{fig:control}).

When the horizon is included, the $uv$-kernel is infinitely extended above floating-point precision due to the interaction of the secondary sidelobes with the horizon edge. While the coordinate distortion caused by the orthogonal projection is an issue (e.g. \citealt{brouw_aperture}), it is not the root cause of this effect. Rather, it is the abrupt transition between beam sensitivity on the sky and an opaque horizon. Regardless of coordinate basis, an abrupt transition creates an infinite response in the Fourier-dual equivalent.  

The FEKO simulation used to model the MWA beam does in fact have this abrupt transition since it uses an infinite plane of soil to model the instrumental response \citep{sutinjo_intrinsic_2013,sokolowski_calibration_2017}. In fact, it is more typical for FEKO simulations to use infinite metallic ground planes, not just infinite soil \citep{bolli_computational_2022}, to reduce complexity at the cost of low elevation accuracy \citep{bolli_Preliminary_2020}. Whether the abrupt edge is a real effect or purely simulation-based is unknown. In our simulations, the beam sensitivity at the horizon is 0.3\% at its largest, highlighting the severity of this effect in $uv$-space.

This infinite extent is detrimental for gridding/degridding analyses, particularly for optimal map making, because it is impossible to represent the instrumental $uv$-kernel well. In fact, we can only perform limited experiments with this kernel since there is no way to include a ground truth (or simulation thereof) as our control. By gridding all foreground sources, we can probe the expected gridding systematics with the analytic convolution kernel, represented out to 30\,$\lambda$ in extent. 

\begin{figure}
\centering
	\includegraphics[width =\columnwidth]{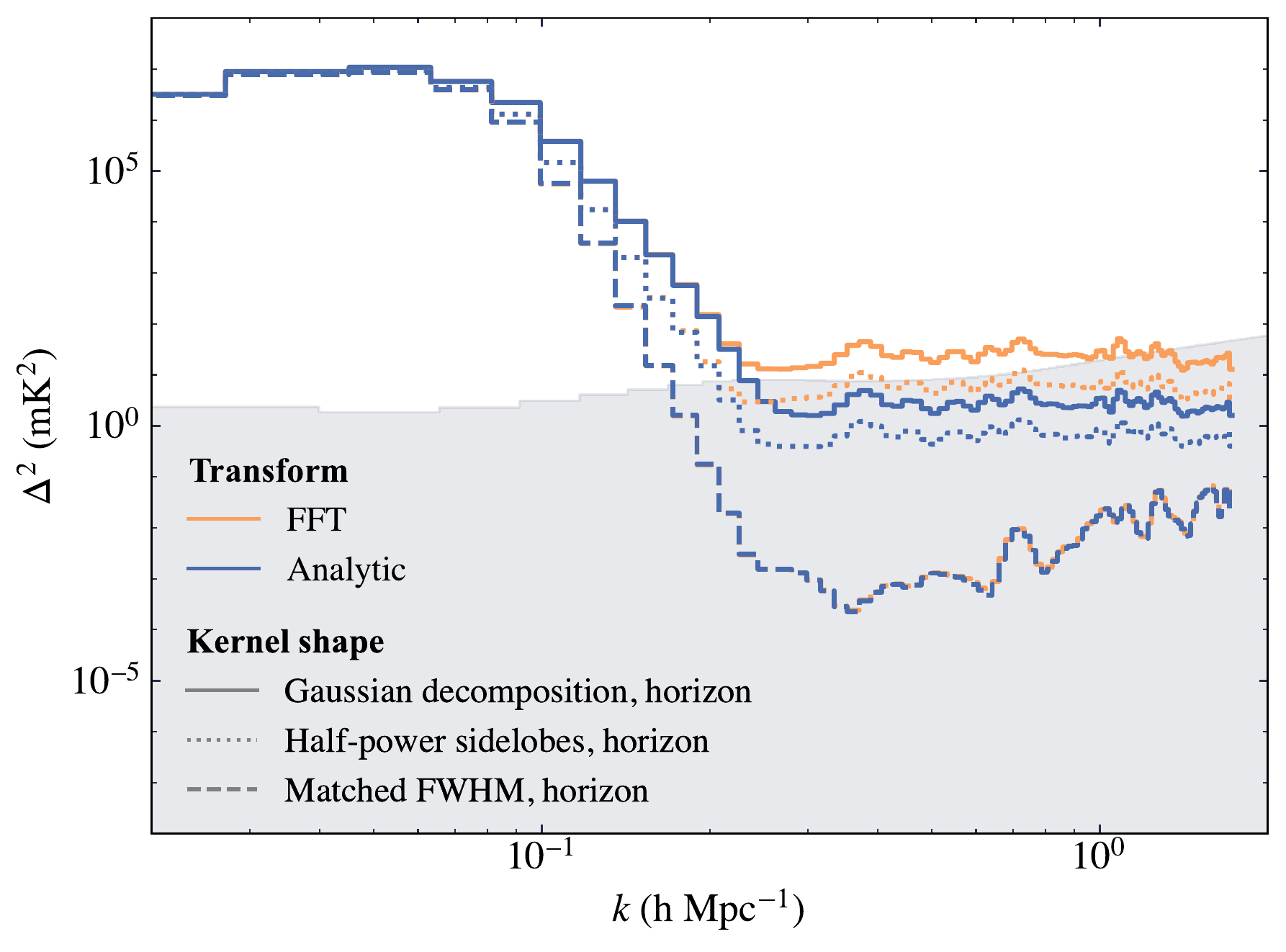}
	\caption{The 1D power spectra as a function of $k$ for various beam shapes with horizons (separated by linestyle) and different transforms (separated by color). The analytic Fourier transforms (blue) of the Gaussian decomposition (solid) and half-power sidelobe (dotted) beam shapes have contamination at high $k$ due to a beam--horizon interaction, and the FFT (orange) of those beam shapes are approximately an order of magnitude worse. The single matched-FWHM Gaussian beam shape (dashed) does not have beam--horizon interactions, and thus traces out the base level of error expected in this simulation. Errors must be within the shaded region in order to theoretically detect the EoR.}
	\label{fig:fft_power}
\end{figure}

The resulting power spectra for the Gaussian decomposition (solid blue, Figure~\ref{fig:fft_power}) shows contamination slightly below the EoR level on high $k$-modes. This is unsurprising given the harsh derivatives at the edge of the kernel, even at +/- 15\,$\lambda$. Whilst this can be mitigated by windowing or tapering the $uv$-kernel to deviate from an instrumental representation, we cannot predict how non-optimal the resulting errors would be. However, given that the kernel is infinite when there is an abrupt transition in beam sensitivity at the horizon, the kernel used in gridding would have to be finite and thus non-optimal anyways. Optimal map making is thus not achievable in the presence of a beam-horizon interaction.

In addition, there is no way to mitigate this contamination for simple calibration without incurring errors reminiscent of those in Figure~\ref{fig:cal_ps}. However, we cannot perform the necessary experiment to constrain requirements of calibration in the presence of the horizon due to the limitations of the in situ simulation framework. 

This contamination can be reduced, however, if the instrumental beam kernel has smaller secondary sidelobes. For the Gaussian decomposition with half-power sidelobes, this high $k$-mode power is now lower (dotted blue, Figure~\ref{fig:fft_power}). If there is little to no beam sensitivity at the horizon, as is the case for the matched FWHM beam, then the contamination traces out the base level of error expected given this simulation resolution (dashed blue, Figure~\ref{fig:fft_power}). 

A further complication arises when FFTs are used to generate the $uv$-kernel. In contrast to the analytic convolution, using FFTs is vastly more efficient and requires significantly less computational load. However, the horizon edge can cause obvious aliasing in the $uv$-domain if the resolution of the image-space beam is not high enough. The base resolution, which is arguably rather small (3,000 pixels in the $l$ direction), causes aliasing contamination as seen in the bottom left panel of Figure~\ref{fig:fft_panel}. Increasing the resolution by 10x (30,000 pixels in the $l$ direction) considerably reduces this contamination (bottom middle panel of Figure~\ref{fig:fft_panel}), and increasing the resolution by 20x (60,000 pixels in the $l$ direction) essentially reduces the extra contamination completely (bottom right panel of Figure~\ref{fig:fft_panel}).

Aliasing in the $uv$-kernel via the FFT can contribute to the power spectra. While easily fixed with image-space filtering, this cannot be done for optimal map making formulisms. The kernel created with the base resolution FFT contributes approximately an order of magnitude higher contamination on high $k$-modes for the beams with sidelobes (solid and dotted orange, Figure~\ref{fig:fft_power}). Of course, when there is little to no beam sensitivity at the horizon, as is the case for the FWHM Gaussian beam, then there is no extra contamination (dashed orange, Figure~\ref{fig:fft_power}). Fortunately, the higher resolution FFT kernels presented in Figure~\ref{fig:fft_panel} do not add significant contamination to the power spectra and are omitted for clarity. The level of contamination expected given a $uv$-kernel of 30\,$\lambda$ can be reached with approximately 5x the image-space resolution. Thus, not only is $uv$-kernel resolution important as explored in \citet{offringa_impact_2019}, but image-space resolution can also contribute to contamination if not properly handled within an optimal map making framework.

Regardless of whether or not the Fourier transform is performed analytically or with a FFT, a beam--horizon interaction leads to a kernel that is infinitely extended. Therefore, it is essentially impossible to represent the instrumental beam well. Thus, if optimal map making and/or simple calibration procedures are to be used for EoR studies, then the instrument must have little to no beam--horizon interaction.


\section{Discussion}

The SKA era promises to push the boundaries of our current understanding of the Universe, and with it, our current analysis methodologies. The Epoch of Reionisation (EoR) science case is an extreme example of required precision and accuracy, in both analysis and measurement, and is thus a benchmark towards the future success of the SKA. Studies have already begun to reevaluate traditional radio astronomy analysis in the context of the EoR. 

In particular, traditional radio astronomy imaging techniques rely on a spreading function, or kernel, in pixelated $uv$-space. Extreme resolution is required to reduce spectral structure caused by the movement of baselines convolved with the $uv$-kernel \citep{offringa_impact_2019}. However, other aspects of the kernel, including its extent, shape, and horizon, also have an important role in reconstructed power spectra analyses. This work extends upon \citet{offringa_impact_2019} to explore these other aspects and how they propagate to calibration, subtraction, and our final measurements, particularly for optimal map making.

To test these effects, we built a semi-analytic framework within a current, working EoR pipeline for the Murchison Widefield Array (MWA). We introduce a Gaussian decomposition of the image-space beam in directional cosine space which allows us to analytically transform and easily modify the gridding shape. The base level of error expected from this framework is 10$^{11}$\,mK$^2$ smaller than the foregrounds, allowing for an unhindered exploration of the aforementioned effects.

Our first experiment quantifies the effect of choosing a gridding kernel shape in optimal map making. This choice is a balance between reducing systematics caused by the edge derivative, increasing the number of pixels required for each gridded visibility, and increasing uncertainties due to non-optimal beam shapes.

For an MWA-style Gaussian decomposition of the beam, the $uv$-kernel must extend out to at least 0.001--0.0001\% of the maximum value in order to measure the EoR at high $k$. Decreasing the sidelobes to half the power does not change this result drastically. However, a kernel with no sidelobes has a much more compact $uv$-kernel, which results in lower high-$k$ systematics for a given kernel size. Even though this is a misrepresentation of the input beam kernel shape, it does not affect high-$k$ modes in the power spectrum, which bodes well for foreground avoidance techniques using gridding. In addition, this only increases uncertainties by roughly 8\% compared to the optimal case. 

Interestingly, all kernel shapes perform poorly at small $k$-modes where foregrounds are prevalent, even the Gaussian decomposition of the beam. The resolution of the Gaussian decomposition of the beam did not match the input, which was semi-analytic in nature, and thus the direct foreground subtraction performed poorly. Given this result and those of \citet{offringa_impact_2019}, simple foreground removal techniques will most likely not reveal the EoR within foreground-dominated regions given current computing resources, and thus more advanced methods will most likely need to be employed.

Our second experiment further supports this point. We use a traditional calibration scheme that calculates a gain parameter for every frequency channel using a linear least squares reduction between input visibilities and a generated model. Even with 100\% of the foregrounds in the model visibilities, the base level of calibration error is orders of magnitude higher than the EoR signal, again due to the difference between the semi-analytic input visibilities and the high-resolution, yet discrete degridded model visibilities. 

We also explore the consequences of misrepresentations of the beam kernel used to generate the input visibilities during the calibration process. An incorrect beam shape drastically increases calibration error such that an EoR measurement is precluded. This is true for both the single Gaussian and the Gaussian decomposition with reduced sidelobes, indicating the importance of correctly modeled sidelobes in calibration. Instrumental beams are vital in calibration procedures to help reduce the base level of error in sky-based, traditional calibration. 

Including the horizon in our framework introduces even more potential challenges. If there is beam sensitivity which interacts with the horizon, even on a small but non-floating-point precision level, then the resulting $uv$-kernel is infinitely extended above machine precision. This cannot be represented well in a discrete, finite analysis, and it will always have high edge derivatives. This generates systematics at about the EoR signal at high $k$-modes in the power spectrum, but this can be easily mitigated by choosing a non-optimal compact kernel. In any case, optimal map making is not an option when there is a beam-horizon interaction because the infinite kernel cannot be represented. However, this poses a potential problem for traditional calibration procedures where the instrumental kernel is not compact.

The inclusion of the horizon in optimal map making frameworks also complicates the use of Fast Fourier Transform (FFT) algorithms, which are discrete and thus cause aliasing in the presence of a large derivative. Compared to the analytic convolution, the FFT of the image-space beam can generate systematics up to an order of magnitude greater. This can be mitigated by increasing the resolution \textit{on the image-space beam}. However, it should be noted that the limitations of the in situ simulation framework cannot probe quantitative requirements in this case, as the true $uv$-kernel is infinitely extended and cannot be represented. 


The simulations presented in the work indicate that gridding is sufficient for foreground avoidance with reconstructed power spectra (Section~\ref{subsec:Beam_kernel_extent}), but that it quickly fails when used in other contexts. Traditional calibration requires degridding resolutions which we cannot obtain (Section~\ref{subsec:calibration_bias}). In addition, inclusion of a horizon for widefield instrumentation creates infinite $uv$-kernels, even when beam sensitivity is as low as 0.3\% at the horizon (Section~\ref{subsec:The_effects_of_the_horizon}). Given this evidence, it is becoming clear that gridding is best used only as a histogramming step, and that using gridding/degridding to calibrate or perform truly optimal map making is relatively unobtainable for EoR reconstructed power spectra studies using widefield instrumentation.

This work, in addition to \citet{offringa_impact_2019}, indicates that advancements beyond traditional radio astronomy imaging methods are needed in order to measure the EoR. Progress over the past decade has started to decouple EoR analysis methodologies from the narrow-field assumptions that built the foundations of radio-astronomy. Whilst it is still possible to use these techniques, namely gridding/degridding, complications arise due to the widefield nature of most EoR interferometers and the spectral requirement for calibration of power spectra analyses, especially within an optimal map making framework. By using an in situ simulation, we explored various aspects of the gridding/degridding kernel in this work, and planned further work will assess mitigation strategies in observational data. 


\section*{Acknowledgements}

The authors would like to thank Cathryn Trott, Bart Pindor, Randall Wayth, Daniel Mitchell, and Miguel Morales for their informative discussions, and Chris Riseley for his advice. N.B. acknowledges the support of the Forrest Research Foundation, under a postdoctoral research fellowship. This research was supported by the Australian Research Council Centre of Excellence for All Sky Astrophysics in 3 Dimensions (ASTRO 3D), through project number CE170100013. This work was supported by resources awarded under Astronomy Australia Ltd’s merit allocation scheme on the OzSTAR national facility at Swinburne University of Technology. OzSTAR is funded by Swinburne University of Technology and the National Collaborative Research Infrastructure Strategy (NCRIS). The International Centre for Radio Astronomy Research (ICRAR) is a Joint Venture of Curtin University and The University of Western Australia, funded by the Western Australian State government. 



\bibliography{sample631}{}
\bibliographystyle{aasjournal}



\end{document}